\documentclass[%
 reprint,
 amsmath,amssymb,
 aps,
]{revtex4-2}

\usepackage{graphicx}
\usepackage{dcolumn}
\usepackage{bm}
\usepackage{float}
\DeclareMathOperator\erf{erf}
\usepackage{placeins}
\begin{document}

\preprint{APS/123-QED}

\title{Multi-Motor Cargo Navigation in Complex Cytoskeletal Networks}

\author{Mason Grieb}
\author{Nimisha Krishnan}
\author{Jennifer L. Ross}%
 \email{corresponding author email: jlross@syr.edu}
\affiliation{%
 Physics Department, Syracuse University}




\date{\today}

\begin{abstract}
The kinesin superfamily of motor proteins is a major driver of anterograde transport of vesicles and organelles within eukaryotic cells via microtubules. Numerous studies have elucidated the step-size, velocities, forces, and navigation ability of kinesins both in reconstituted systems and in live cells. Outside of cells, the kinesin-based transport is physically regulated and can be controlled by obstacles or defects in the path, or the interaction between several motors on the same cargo. To explore the physical control parameters on kinesin-driven transport, we created complex microtubule networks in vitro to test how kinesin cargoes made from quantum dots with one to 10 kinesin motors attached are able to navigate the network. We find that many motors on the quantum dot significantly alter distance walked, time spent bound,  the average speed, and the tortuosity of the cargo. We also find that the average mesh size of the microtubule network affects the end-to-end distance of the motion, the run time, average speed and tortuosity of cargoes. Thus, both motor number and network density are physical aspects that regulate where cargoes traverse in space and time.

\end{abstract}

\maketitle


\section{INTRODUCTION}

The cell interior is a complex physical space with objects ranging from the nanoscale to the millimeter scale, depending on the cell type. For instance, yeast cells are tiny, on the order of 4 $\mu$m in diameter, and spherical, while neurons in adult humans can be a meter in length and be almost 2.5 meters long inside a giraffe's neck. An essential aspect of life and the maintenance of living systems is the ability to transport objects in this complex environment, often along relatively long distances. Large cells, including most animal cells, are large enough to require transport machinery to move objects small to large (i.e., molecular to organelle-sized) to where they need to be in space and time \cite{sheetz1989mechanism}. The main apparatus for this long-distance transport is the microtubule cytoskeletal system and the enzymatic motor proteins that can bind and walk along microtubules \cite{vale2003molecular, hirokawa1998kinesin}. 

Microtubules are hollow, cylindrical structures made from the non-covalent binding of tubulin dimers. They have an intrinsic asymmetry in their structure so that the two ends display different dynamics  with the ``plus end'' topped with a beta tubulin dimer and the ``minus-end'' capped with an alpha tubulin dimer \cite{ross2008cargo}. The motor proteins that deliver cargoes in the cell are from two main families: kinesin and dynein, which generally move on microtubules by walking in an hand-over-hand manner to the plus-end or minus-end, respectively \cite{ross2008cargo}. The overall organization of microtubules in cells is part of the control mechanism to guide cargoes to where they are needed spatio-temporally. While much of recent structural biology has focused on the vesicular cargo adapters for the motors, which dictate how the motors bind to cargoes \cite{siddiqui2019ptpn21,verhey2001cargo,yang2005kinesin}, there are still open questions about the physical nature of intracellular transport. Specifically, how does the filament organization control the when and where of cargo transport? 

\begin{figure}[hbt!]
\centering
\includegraphics[scale=0.78]{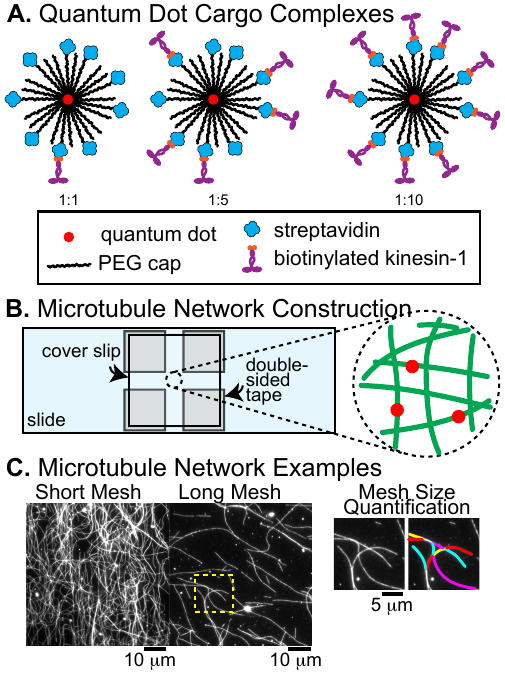}
\caption{{\bf Experimental approach for multi-motor cargoes in complex microtubule networks.}
A) Multi-motor cargoes are composed of CdSe quantum dots cpped with poly-ethylene glycol (PEG) polymers and covalently-bound streptavidin. Biotinylated, truncated kinesin-1 motors are incubated at various concentrations to have quantum dot-to-motor stoichiometries of 1:1, 1:5, and 1:10. B) The experimental chamber is composed of a slide and coverglass with four pieces of double stick tape to create a crossed flow path. The central region is the location where large networks are formed. The networks have crossing filaments at various densities, depending on the amount of microtubules flowed through and adhered to the cover glass. C) Example images of microtubule networks with low mesh size (high density) and high mesh size (low density). The mesh size is quantified by estimating the distance between intersections.}
\label{methodsfigure}
\end{figure}

To explore the effects of increasingly complex physical environment on multi-motor cargo transport, we take a bottom-up approach to create dense microtubule networks. Specifically, we use wholly purified tubulin to create microtubules and organize them into crossing filament networks with different mesh sizes. Next, we create synthetic cargoes with increasing amounts of motors bound to them using highly fluorescent quantum dots with 1, 5, or 10 motors bound (Fig. \ref{methodsfigure}). Using this system, we find interesting effects of both the motor number and the mesh size as these cargoes traverse the network. Specifically, we find that the distance the cargoes move depends on the number of motors attached to the cargo, which is not surprising. More surprisingly, we find that the speed of the cargoes depends on motor number, which should only happen if the cargoes are moving in an environment with high viscosity or drag. In examining the effect of the network density, we find that the network itself is acting as the drag on the cargoes. These results recapitulate prior results from live cells that demonstrated a speed-dependence on the number of motors \cite{efremov2014delineating,kural2005kinesin,macosko2008fewer,gagliano2010kinesin}, and show that the physical nature of the network can influence the transport of cargoes with single or multiple motors.

\section{MATERIALS AND METHODS}
\subsection{Reagents and Proteins}
Chemical reagents are purchased from Sigma, unless otherwise noted. Tubulin dimers purified from porcine brains are purchased from Cytoskeleton. Biotinylated kinesin-1 (b-K401) is expressed from plasmid DNA (AddGene, pWC2, \#15960) in E.coli and purified using previously published protocols \cite{pierce199814,conway2014measuring}. Briefly, bacteria are grown to OD of 0.6 and IPTG is added overnight to express protein. Bacteria are lysed and the lysate containing the protein of interest is clarified and then incubated with nickel-ion beads. Beads are loaded into a gravity column and protein is eluted using imidizole. Fractions are collected and protein is tested using protein gel electrophoresis to quantify purity and concentration. Quantum dots with an emission wavelength of 640 nm and capped with poly-ethylene glycol (PEG) and covalently labeled with streptavidin are purchased (Life Technologies) and stored at 4$^{\circ}$C (Fig. \ref{methodsfigure}A).

\subsection{Quantum Dot Cargo Preparation}
Quantum dot cargoes are created by mixing purified b-K401 with 10 $\mu$M quantum dots. The b-K401 is mixed with the cargo at molar ratios of 1:1, 1:5, and 1:10 (Fig. \ref{methodsfigure}A) and are incubated for at least 30 minutes to allow time for the complexes to form \cite{conway2012motor,krishnan2023effects}. As we are relying on diffusion to distribute motors onto the cargo there is a chance that not all cargo molecules have exactly 1, 5, or 10 motors, but should have these numbers on average. The cargoes are used to create an active mix by diluting the quantum dot-kinesin mix by a factor of 15 into a solution with 13 $\mu$M Taxol, oxygen scavenging system (12 mg/ml glucose oxidase and 5.9 mg/ml catalase suspended in deionized water), 20 mg/ml glucose, 67 mM Dithiothreitol (DTT), 1 mM adenosine triphosphate (ATP), all in PEM-80 buffer (80 mM PIPES pH 6.9, 2 mM MgCl2 and 0.5 mM EGTA). The entirety of the mixture is flowed into the chamber from the bottom entrance before starting video recording.

\subsection{Microtubule Network Preparation}
Microtubules are polymerized from 5 mg/ml tubulin in PEM-80 buffer by mixing 90\% unlabeled tubulin with 10\% fluorescently-labeled tubulin (Hilyte 488) and 1 mM guanosine triphosphate (GTP) and incubating for 20 minutes at 37$^{\circ}$C. To stabilize the filaments, Taxol is added to a final concentration of 8.7 $\mu$M and incubated a second time for 20 minutes at 37$^{\circ}$C. Stable microtubules are kept at room temperature on the lab bench at high concentration for up to two weeks. Microtubules are diluted by 1:100 prior to flowing into the chamber.

All of our data is taken within crossed flow path chambers constructed by placing four pieces of double sided tape onto a slide then a cover slip was placed on the tape leaving four entrances to the chamber in a cross configuration (Fig. \ref{methodsfigure}B), as previously described \cite{ross2008kinesin,krishnan2023effects}. Cover slips are treated with hydrophobic silane (PlusOne Repel Silane SE, Cytiva) using the process previously described \cite{dixit2008differential,conway2012motor}. To bind the microtubules to the surface, 15 $\mu$l of 10$\%$ alpha tubulin antibody (YL1/2, Millipore) dissolved in PEM-80 is added into the chamber. The chamber is then allowed to sit for 5 minutes, cover slip side down, to allow the antibody to bind to the surfaces. Next 8 $\mu$l of 5$\%$ F127 is flowed into the chamber in one direction, then another 8 $\mu$l is flowed into the chamber in the perpendicular direction, and the chamber is then allowed to sit for another 5 minutes. Next, we flow a wash mixture consisting of 0.05\% F127 in PEM-80, again using  8 $\mu$l in each perpendicular direction. Next 8 $\mu$l of a 1:100 diluted microtubule mixture is flowed into the chamber in first one direction and allowed to sit for 2 minutes. After this, we flow another 8 $\mu$l of wash from the same direction and leave the slide for another 3 minutes. After the first direction of microtubules is deposited, the process is repeated in the perpendicular direction. This leads to a microtubule network that contains a roughly even ratio of microtubules in each direction, creating a dense, crossed network in the center of the chamber (Fig. \ref{methodsfigure}B,C). 

\subsection{Imaging and Image Analysis}
The quantum dot cargoes and microtubule networks are imaged with total internal reflection fluorescence (TIRF) microscopy using a Nikon Eclipse Ti2 microscope with an iLas (GATACA systems) attachment for laser illumination using the ring TIRF modality for a smooth illumination region. Both microtubules and quantum dots are illuminated with lasers (488 nm and 638 nm) and fluorescence is recorded simultaneously onto two Prime BSI scientific CMOS cameras (Photometrics) with an ROI of 1192 x 1192 pixels using Nikon Elements acquisition software. The green emission (microtubules) is split first and imaged to a camera mounted to the back of the system. The red emission (quantum dots) is passed through to the camera on the left of the microscope. Data are recorded as 2 minute videos with 1 second between each frame and an exposure time of 100 ms. Movies are saved as stacks of tiffs with metadata as .nd2 files. Multiple movies are recorded using the same chamber for up to 30 minutes, which is enough time to ensure that the motors did not run out of ATP. The regions selected to record were chosen based on the size of the microtubule mesh in the region to acquire data at a variety of mesh sizes. 

Images are analyzed using FIJI/ImageJ. Each large movie was divided into quadrants (596 x 596 pixels) because the mesh size of the network is different in different areas in the image. This allowed for the average mesh size to better represent the region. Microtubule network density is analyzed using the skeletonize plugin in FIJI \cite{arganda20103d}, as previously described \cite{krishnan2023effects}. A single grayscale image of the microtubule (green) channel is saved as a .tif image which is then binarized using the threshold function. The skeletonized feature is used to measure an average distance between intersections allowing us to quantify the mesh size. Branches less than 0.5 $\mu$m were deleted before averaging because the optical resolution of the imaging is 0.5 $\mu$m. We find that the skeletonize feature does not capture the mesh perfectly and frequently adds short imaginary branches to regions with high brightness or background aggregates, this is reduced by running the ``close'' function in FIJI on the binary image before skeletonizing. An example of short mesh length (high density network) and long mesh length (low density network) is shown in figure \ref{methodsfigure}C.

Using the quantum dot channel (far red), trajectories are determined with manual tracking using FIJI/ImageJ Manual Tracking plug-in. We find that automated tracking algorithms we tried disconnected trajectories compared to the manual tracking. The resulting .csv files of the trajectories spatial location (x,y) over time were boxcar averaged over three frames to smooth the data and remove spurious fluctuations perpendicular to the microtubule track that did not represent the actual trajectory. Data was further analyzed using Python to extract the contour length, end-to-end displacement, and association time for each track. The 1:1 data set consists of 36 analyzed quadrants from 9 different videos, with a total of 881 trajectories were measured. The 1:5 data set consists of 11 analyzed quadrants from 7 different videos with a total of 270  trajectories measured. The 1:10 data set consists of 15 analyzed quadrants from 8 different videos with a total of 326 trajectories measured. 

\begin{figure}[hbt!]
\centering
\includegraphics[scale=0.9]{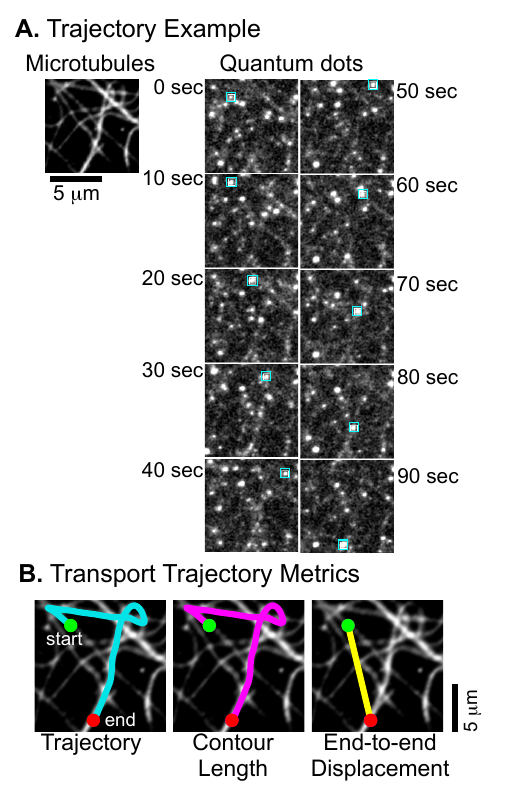}
\caption{{\bf Transport trajectory metrics.}
A) Example of microtubule network (left) with frames from a recording over time with 10 second intervals showing the trajectory of a quantum dot cargo (right). The cargo of interest is boxed in cyan and moves from the upper left to the right and then diagonally to the lower left. Scale bar is 5 $\mu$m. Example from 1:10 data set with mesh size of 2.3 $\mu$m. B) Trace of the cargo trajectory (cyan) for the quantum dot from part (A) from the starting position (green dot) to the ending position (red dot). The contour length (magenta) is the total length of the trajectory from start to finish. The end-to-end displacement (yellow) is the length between the start and end locations of the trajectory.}
\label{Example}
\end{figure}

\section{RESULTS AND DISCUSSION}

\subsection{Transport Distance Depends on Motor Number}
Quantum dot cargoes incubated with increasing molar ratios of kinesin-1 were introduced into chambers with crossed microtubule networks of variable mesh sizes (Fig. \ref{methodsfigure}). Image data was recorded (Fig. \ref{Example}A) and images were analyzed to quantify the trajectories (Fig. \ref{Example}B). From the trajectories, we measure the the total length of the trajectory, called the contour length, and the end-to-end displacement of the trajectory (Fig. \ref{Example}B). 

\begin{figure}[hbt!]
\centering
\includegraphics[scale=0.9]{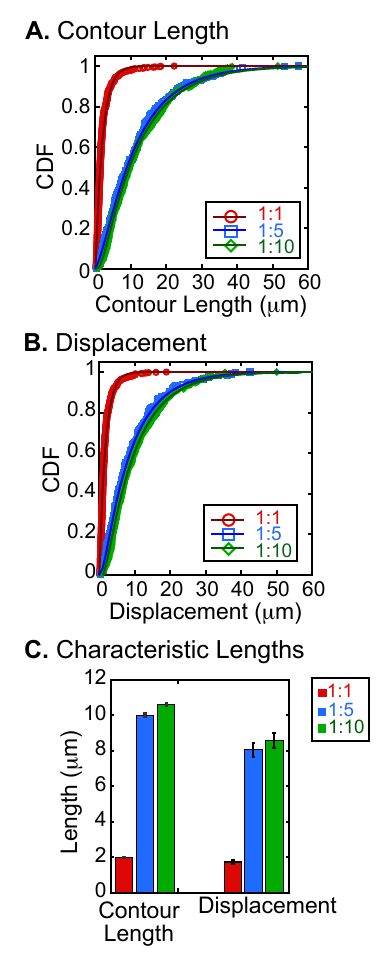}
\caption{{\bf Quantification of transport metrics.}
A) Cumulative distribution function plots of the contour lengths of the trajectories for 1:1 (red circles), 1:5 (blue squares), and 1:10 (green diamonds) are fit to exponential decays with equation \ref{eqn:singleExp} for 1:1 and equation \ref{eqn:doubleExp} for 1:5 and 1:10, same color lines. Best fit parameters given in Appendix Table \ref{tab:Fig2Ci}. B) Cumulative distribution function plots of the end-to-end displacement of the trajectories for 1:1 (red circles), 1:5 (blue squares), and 1:10 (green diamonds) are fit to exponential decays with equation \ref{eqn:singleExp} for 1:1 and equation \ref{eqn:doubleExp} for 1:5 and 1:10, same color lines. Best fit parameters given in Appendix Table \ref{tab:Fig2Cii}. C) The characteristic lengths determined from the best fits for the contour length data and the displacement data for 1:1 (red bars), 1:5 (blue bars), and 1:10 (green bars). Error bars represent the uncertainty of the fit. The statistical differences between the data sets can be found in Appendix Tables \ref{tab:KSFig2A} and \ref{tab:KSFig2B}.}
\label{ContourLength-Displacement}
\end{figure}

We find that the motor number dictates the transport distances for the cargoes (Fig. \ref{ContourLength-Displacement}). Specifically, cargoes with an average of one motor had a far shorter total travel distance and end-to-end displacement compared to cargoes with 5 or 10 motors (Fig. \ref{ContourLength-Displacement}A,B). We plot the data as cumulative distribution functions (CDFs) because every data point is weighted evenly and there is no binning needed. Theoretically, single motors should have an exponentially distributed travel distance, which has an exponentially decaying rise as the CDF: 
  \begin{equation}
        y=1-e^{(-x/\lambda)}
        \label{eqn:singleExp}
    \end{equation}
where lambda is the characteristic decay length of the distribution, also equal to the mean of the distribution. The exponential decay is indicative of a process that has a constant probability per unit time. In this case, for a single motor stepping, there should be a constant probability that both motor heads will become unbound during any step. That results in the single exponential decay \cite{block1990bead,vale1996direct}.

Equation \ref{eqn:singleExp} fits well for the cargoes with one motor (1:1), but not as well for those with five or ten motors (Appendix Table \ref{tab:Fig2Ci}). Instead, these distributions fit better with two exponentials that are each weighted with an amplitude in front of the exponential:
  \begin{equation}
        y=1-Ae^{(-x/\lambda_1)}-(1-A)e^{(-x/\lambda_2)}
        \label{eqn:doubleExp}
    \end{equation}
where $A$ and $1-A$ are the amplitudes of the two exponentials and $\lambda_1$ and $\lambda_2$ are the characteristic decay lengths. When the double exponential was tried for the 1:1 data, the fit gave two exponential with the same decay length within the uncertainty of the fit parameters, implying that two exponentials were not necessary. For the 1:5 and 1:10 data sets, the exponential with the larger amplitudes were those with much longer decay lengths (see Appendix Table \ref{tab:Fig2Ci}). These lengths represent the characteristic contour length and end-to-end displacement (Fig. \ref{ContourLength-Displacement}C). 

The second decay lengths for the 1:5 and 1:10 cargoes were short, and the amplitude was negative, which is indicative of a process that grows exponentially at short time (see Appendix Table \ref{tab:Fig2Ci}). This implies that for multi-motor cargoes, very short runs are suppressed, which has been seen for multi-motor complexes previously \cite{feng2018motor,arpaug2019motor}. It has been theorized that this is due to rapid re-attachment of the second motor, which is in close proximity \cite{arpaug2019motor}. Indeed, for a quantum dot, when 5-10 motors are attached, they are likely quite close and possibly able to attach at the same time, suppressing very short run lengths.

\subsection{Transport duration, speed, and tortuosity depend on motor number}
\begin{figure}[!]
\centering
\includegraphics[scale=0.9]{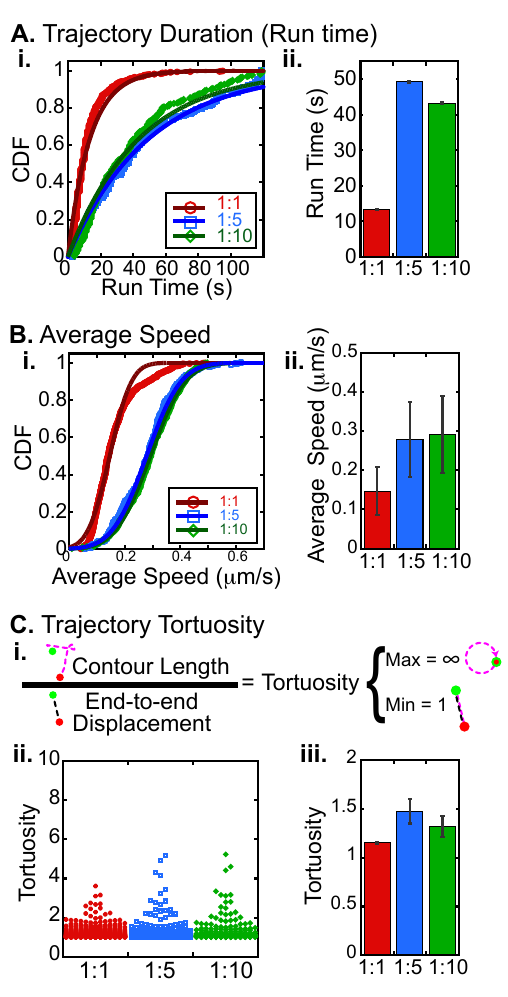}
\caption{{\bf Motor numbers affect duration, average speed, and tortuosity of trajectories.}
A) Quantification of the trajectory duration (run time). (i) CDF plot of run time for 1:1 (red circles), 1:5 (blue squares), and 1:10 (green diamonds) quantum dot:motor ratios. Fits use equation \ref{eqn:singleExp}. Best fit parameters given in Appendix Table \ref{tab:Fig3A}. (ii) Average run times determined from the best fits for 1:1 (red bar), 1:5 (blue bar), and 1:10 (green bar) with error bars denoting the uncertainty in the fit parameters. B) Quantification of the trajectory average speed. (i) CDF plot of average speed for 1:1 (red circles), 1:5 (blue squares), and 1:10 (green diamonds) quantum dot:motor ratios. Fits use equation \ref{eqn:Gaussian} to determine the mean of the distribution. (ii) The mean average speed for 1:1 (red bar), 1:5 (blue bar), and 1:10 (green bar) with error bars denoting the variance, determined from the fit equation. C) Quantification of the tortuosity. (i) Tortuosity is the ratio of the contour length to the end-to-end displacement. The range is 1 to infinity. (ii) Dot distribution for tortuosity values for each trajectory for 1:1 (red circles), 1:5 (blue squares), and 1:10 (green diamonds) quantum dot:motor ratios.  (ii) Calculated average tortuosity for 1:1 (red bar), 1:5 (blue bar), and 1:10 (green bar) data with error bar representing the calculated standard deviation of the distribution.}
\label{Duration-Speed-Tortuosity}
\end{figure}

Using the same trajectories quantified for the transport distances, we can also measure the duration of the motility. We find that the motor number dictates the duration of the transport, or the run time, for the cargoes (Fig. \ref{Duration-Speed-Tortuosity}A). As expected, more motors enable longer time spent bound to the microtubules because a cargo with more motors will have more local binding sites for the microtubule and allow for simultaneous engagement by multiple motors on the same cargo or rapid reattachment before the cargo diffuses away. 

The run time distributions fit to single exponential curves (Eqn. \ref{eqn:singleExp}) for all data sets (Fig. \ref{Duration-Speed-Tortuosity}Ai). When a sum of two exponentials was tried, the fit parameters were identical, implying only one exponential decay was needed (Appendix Table \ref{tab:Fig3A}). The characteristic run time from the fit equation shows that having 5 or 10 motors increased the association time by almost a factor of five, but there was not much of a difference between 5 and 10 motors (Fig. \ref{Duration-Speed-Tortuosity}Aii). Using the Kolmogorov-Smirnov statistical test (KS Test), we find that the probability that these two distributions are different to be 22\%, so we conclude that the attachment duration for 5 and 10 motors are identical (Appendix Table \ref{tab:KSFig3A}).

Using the contour length and the run time of each individual trajectory, we can take the quotient to quantify the average speed of the cargoes. Interestingly, we find that the average speed also depends on the number of motors attached to the cargoes. Plotting the distribution as a cumulative distribution, the data is clearly Gaussian (Fig. \ref{Duration-Speed-Tortuosity}Bi), and not an exponential as the duration or trajectory lengths were. We fit the CDF to a Gaussian in the form:
\begin{equation}
        y=\frac{1}{2} \left(1-\erf{\left(\frac{x-\mu}{\sigma \sqrt{2}}\right)}\right)
        \label{eqn:Gaussian}
    \end{equation}
where $\erf$ is the error function, $\mu$ is the mean of the distribution, and $\sigma$ is the standard deviation, which is the square-root of the variance (Appendix Table \ref{tab:Fig3B}). The average speed for the 1:1 cargoes is significant lower than for quantum dots with 5 or 10 motors. The fit mean, $\mu$ is plotted comparing between the 1:1, 1:5, and 1:10 data (Fig. \ref{Duration-Speed-Tortuosity}Bii). The error in the fit parameters are small, but instead we display the standard deviation as the error bars. The statistical differences are given in Appendix Table \ref{tab:KSFig3B}.

It might be surprising for cargoes with more motors to travel faster than those with fewer motors. On average, we would anticipate that all cargoes should have the same average velocity or even cargoes with more motors should be slower because they should only go as fast as the slowest motor engaged. Several prior studies both in live cells and in vitro have demonstrated that multiple motors can move faster if there is significant drag on the system \cite{efremov2014delineating,kural2005kinesin,macosko2008fewer,gagliano2010kinesin,furuta2013measuring,tjioe2019multiple,levi2006organelle,hill2004fast}. 

Another parameter of interest when transporting in complex networks is the tortuosity, which is defined as the contour length divided by the end-to-end displacement (Fig. \ref{Duration-Speed-Tortuosity}Ci). Given this definition, the tortuosity can be at minimum equal to 1 for a trajectory that is perfectly straight and at maximum infinity for a trajectory that starts and ends at the same position while moving a long distance in between. 

Using the individual trajectory contour lengths and trajectories, we determined the tortuosity to find that cargoes with single motors have the smallest tortuosity, close to one. This makes sense because single motors are known to move relatively short distances, on the order of 2 $\mu$m, and are more likely to dissociate due to the probability of dissociating during stepping or from approaching an intersection \cite{ross2008kinesin,krishnan2023effects}. The tortuosity for the 5 and 10 motor cargoes are both longer than single motors, but not significantly so (Appendix Table \ref{tab:KSFig3C}). For instance, the average calculated tortuosity for the 1:5 cargoes is only about 50\% larger than single motors (Fig. \ref{Duration-Speed-Tortuosity}Ciii). Using the KS Test, we determine that the probability that the :1 and 1:5 cargoes have distinct distributions of tortuosity is only 8\% (Appendix Table \ref{tab:KSFig3C}).

\subsection{Cargoes with more motors cross and turn at intersections more frequently}
\begin{figure}[hbt!]
\centering
\includegraphics[scale=0.9]{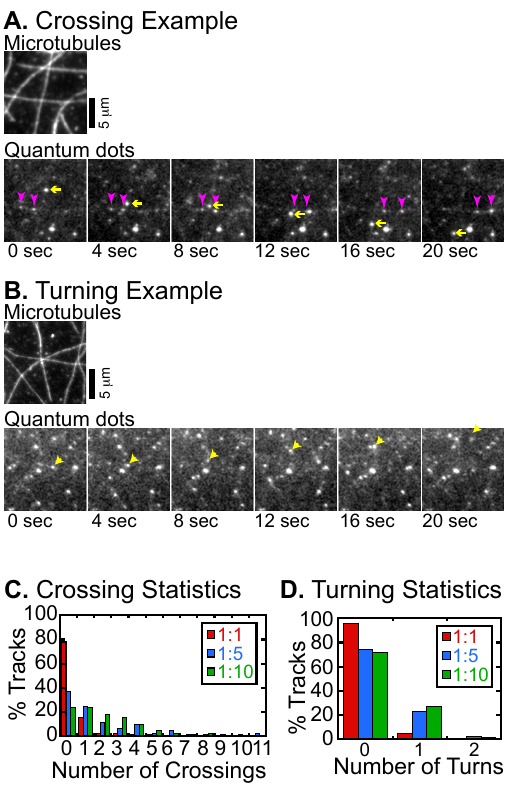}
\caption{{\bf Cargoes with more motors turn and cross at intersections.}
A) Example of trajectories that pass intersections. (Top) Image of microtubule network. (Bottom) Time series with four second intervals showing several cargoes approaching the intersection and crossing including one that starts on a vertical microtubule above the intersection and travels down and past the intersection (horizontal, yellow arrow) and two cargoes on the horizontal microtubule that travel to the right and pass the intersection (vertical magenta arrowheads). Scale bar is 5 $\mu$m. B) Example of trajectory that turns at an intersection. (Top) Image of microtubule network. (Bottom) Time series with four second intervals showing cargo that turns at the intersection starting on the horizontal microtubule, traveling to the left and then turning onto the vertical microtubule travelling up (yellow arrowhead). Scale bar is 5 $\mu$m. C) Quantification of crossing statistics for cargoes with 1:1 (red bars), 1:5 (blue bars), and 1:10 (green bars) cargo:motor ratios. D) Quantification of turning statistics for cargoes with 1:1 (red bars), 1:5 (blue bars), and 1:10 (green bars) cargo:motor ratios.}
\label{Crossing-Turning}
\end{figure}

The system we created has two kinds of complexity - the cargoes have multiple motors and the network has multiple intersecting filaments. The intersections of the filaments are an opportunity for the cargoes to turn and change direction, but they can also act as obstacles to forward motion\cite{osunbayo2015cargo}. Using particle tracking, we examined the ability of the cargoes to cross at intersections as well as turn (Fig. \ref{Crossing-Turning}A,B). 

We find that the cargoes with a single motor were least likely to cross or turn at intersections and most likely to dissociate at the intersection, while multi-motor cargoes with 5 or 10 motors were able to both cross and turn at intersections (Fig. \ref{Crossing-Turning}C,D). Specifically, almost 80\% of 1:1 cargoes dissociated at intersections, implying that the intersection was a barrier to forward motion. Of the remaining 20\% that interacted with the intersection and remained bound, over 95\% went straight past the intersection and only 5\% were able to turn. Cargoes with more motors were much more likely to turn at the intersection, over 20\% of the time, which is likely due to the fact that another motor on the same cargo was probably capable of engaging with the crossing filament. These probabilities are similar to those reported previously for single motors and large beads coated with multiple motors \cite{ross2008kinesin}.   

\FloatBarrier

\subsection{Multi-motor cargoes trajectories depend on the network organization}
Cargoes with multiple motors are examined in complex microtubule networks with a variety of distances between intersections, characterized by the mesh size (Fig. \ref{methodsfigure}C). Above, the data for all different mesh sizes was compiled, but the average mesh size of a local area can be determined to investigate if there are any effects due to the mesh size \cite{nelson2009random}. We use regions of networks with similar characteristic mesh sizes and quantify the transport characteristics for cargoes within that region (number of trajectories for each region are given in Appendix Table \ref{tab:Numb-Mesh}). We will also restrict our analysis to examining the 1:5 and 1:10 motors, since another recent study already showed the effects of mesh size on single motors \cite{krishnan2023effects}, and the single motor-driven cargoes have very different results compared to the multi-motor cargoes. Further, we will combine the data for cargoes with 5 motors and 10 motors together, since these data are quantitatively the same in all their transport metrics (Figs. \ref{ContourLength-Displacement}-\ref{Duration-Speed-Tortuosity}). 

First, we examine the average contour lengths and displacement for trajectories in regions with mesh sizes from 0.5 $\mu$m to above 3 $\mu$m. The multi-motor cargoes show that the contour length, or total run length, is independent of the mesh size of the network. We try to fit the data to a line, and the slope of the best fit was $-0.2 \pm 0.9$, which is indistinguishable from a slope of zero. We represent the fit as a horizontal line at the average value of all the data, specifically 12.0 $\mu$m. The median was also 12.0 $\mu$m and the standard deviation was 2.7 $\mu$m.  

It is surprising that the contour length of the trajectories does not depend on mesh size because it indicates that the cargoes are able to take the same number of steps regardless of the network construction or the presence of intersections. Indeed, figure \ref{ContourLength-Displacement} shows that the multi-motor cargoes are able to cross and turn at intersections.  Importantly, this is completely different from the results for single motors, which dissociate and thus end their runs as the mesh size decreases \cite{krishnan2023effects}. Thus, for multi-motor cargoes, there is truly a constant probability of stepping and a fixed probability of releasing at any given step that only depends on there being multiple motors, but does not depend on the filament organization.  

\begin{figure}[h!]
\centering
\includegraphics[scale=0.9]{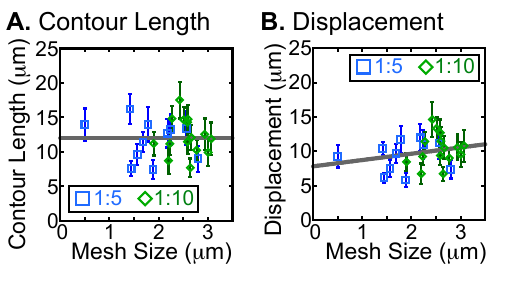}
\caption{{\bf Network mesh size affects the trajectory displacement but not the contour length.}
A) Scatter plot of the average contour length for quantum dots with 5 motors (blue squares) and 10 motors (green diamonds) as a function of the local network mesh size. Error bars represent standard error of the mean for the measurements made in the same region with the average network mesh size. Line represents the average of all the measurements shown, since the best fit line slope was zero. B) Scatter plot of the average end-to-end displacement length for quantum dots with 5 motors (blue squares) and 10 motors (green diamonds) as a function of the local network mesh size. Error bars represent standard error of the mean for the measurements made in the same region with the average network mesh size. The fit equation is a line with fit parameters given in Appendix Table \ref{tab:Fig5B}. }
\label{VMeshLengths}
\end{figure}

Unlike the contour length, the end-to-end displacement length does depend on the mesh size with a linear dependence (Fig. \ref{VMeshLengths}B). The best fit line has a positive slope of $0.9 \pm 0.7$ indicating that as the distance between intersections grows, so does the displacement (fit parameters in Appendix Table \ref{tab:Fig5B}). This result is reasonable because short mesh sizes have closer intersections and multi-motor cargoes are able to turn at intersections  with higher probability. Turning at the intersection reduces the displacement, and is thus more likely when the mesh size is small.

\begin{figure}[t!]
\centering
\includegraphics[scale=0.9]{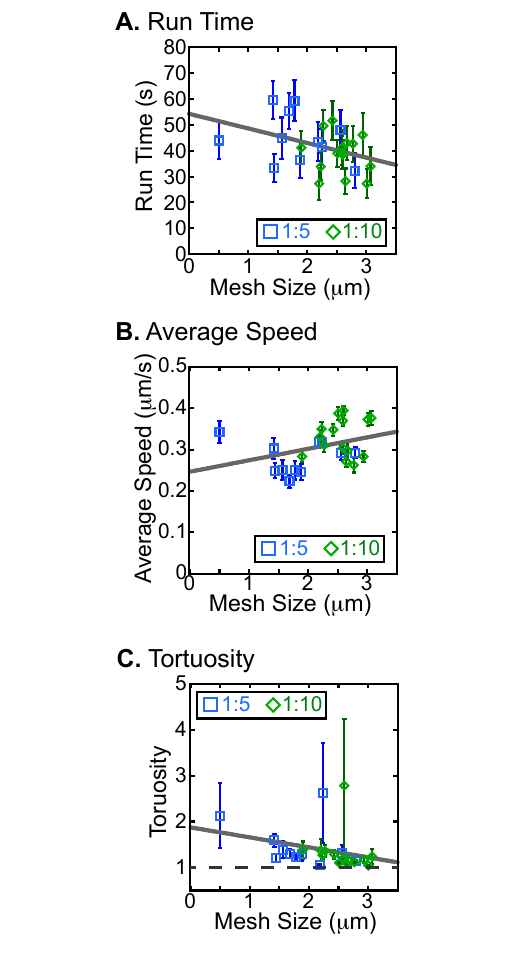}
\caption{{\bf Network mesh size affects the trajectory duration, average speed, and tortuosity.}
A) Scatter plot of the average trajectory run time for quantum dots with 5 motors (blue squares) and 10 motors (green diamonds) as a function of the local network mesh size. Error bars represent standard error of the mean for the measurements made in the same region with the average network mesh size. B) Scatter plot of the average speed for quantum dots with 5 motors (blue squares) and 10 motors (green diamonds) as a function of the local network mesh size. Error bars represent standard error of the mean for the measurements made in the same region with the average network mesh size. C) Scatter plot of the average tortuosity for quantum dots with 5 motors (blue squares) and 10 motors (green diamonds) as a function of the local network mesh size. Error bars represent standard error of the mean for the measurements made in the same region with the average network mesh size. All line fits given in Appendix Table \ref{tab:Fig6}. The dashed line is the minimum value of  tortuosity, which is one. }
\label{VMeshTimeSpeedTort}
\end{figure}

Interestingly, the total duration of the trajectory, or run time, also seems to depend on the network mesh size (Fig. \ref{VMeshTimeSpeedTort}A). Specifically, the run time decreases linearly as the mesh size increases (fit parameters given in Table \ref{tab:Fig6}). Larger mesh sizes mean a longer distance between intersections, which should have an increased path free from crossing filaments that serve as obstacles for single motors. The expectation might have been that these longer transit times on a single filament should have longer association. But, there could be another explanation. Given the small size of the quantum dot, it is likely that during the transit between intersections, only a single motor is transporting. Thus, the dissociation kinetics will be controlled by the dissociation kinetics of a single motor at that time. At intersections, the cargoes have the opportunity to interact with multiple motors on multiple filaments, which will increase the association time. Thus, it is possible that networks with higher density enable the cargoes to stay associated longer because there is more time spent engaging multiple motors on multiple filaments.

Given the constant contour length and decreasing association time with network mesh size, it is not surprising that the average speed of the multi-motor cargoes, which is the quotient of these quantities, also depends on the mesh size (Fig. \ref{VMeshTimeSpeedTort}B). Specifically, cargoes traveling on denser mesh networks (shorter mesh sizes) move slower than on more spares networks (best fit parameters given in Appendix Table \ref{tab:Fig6}). As described above, the denser network is likely serving as large pool of binding sites for multiple motors, which are also acting as a drag to slow the cargoes. This also complements the idea that the network is creating drag on the cargoes, which is why cargoes with more motors are able to move faster than those with single motors (Fig. \ref{Duration-Speed-Tortuosity}B). 

The toruosity also depends on the mesh size with longer mesh size networks having lower tortuosities closer to one and short mesh sizes with higher tortuosities (Fig. \ref{VMeshTimeSpeedTort}C). This makes sense, since lower mesh sizes present more intersections, which are opportunities to turn. More turns lead to shorter end-to-end displacements and thus higher tortuosities. The data can be fit with a line, but there were several networks that had anomalously very high tortuosities, which may be due to one or two cargoes that have high tortuosities. We can remove these anomalously high points (Appendix Figure \ref{masked}), and the fit becomes better, still with a negative slope. The fit parameters for both data sets are given in the Appendix Table \ref{tab:Fig6}.

\section{CONCLUSIONS}

Transport properties of macromolecular cargoes are important biologically and intrinsically interesting to study from a physics perspective. Here, we show that both the increased complexity of the cargoes, specifically the number of motors on each cargo, and the complexity of the network the cargoes are navigating have impacts on the transport properties. As expected, increasing the number of motors increases the distance the cargo moves as the time associated. This is reasonable because the probability that two motors dissociate at the same time is significantly lower than for a single motor. Multiple motors also allow the cargoes to switch at intersections, increasing the tortuosity of the motion. It is surprising that multiple motors increase the speed of the transport, implying that the heavily crossed network creates a drag that multiple motors can overcome. This effect is clear when the data is plotted against the regional mesh size. The speed increases as the distance between intersections increases, implying it is the mesh causing the slow down. This is likely due to multiple motors being engaged simultaneously, possibly at intersecting microtubules. Future work with different types of kinesin motors with different inherent transport properties or using cytoplasmic dynein would continue to explore the physical effects on motor-cargo transport.

\begin{acknowledgments}
This work was conceived of by NK and MG with guidance from JLR. MG did the majority of the experiments and analysis. This work was partially supported by funding from Syracuse University, the National Science Foundation grants NSF BIO-2134215, NSF DMREF-2118403, and National Institutes of Health grant 1R15GM141722 to JLR. We thank the Ross Lab members for their help with this project. 
\end{acknowledgments}

\appendix

\section{Statistical analysis and fit data}\label{appendixA}
This appendix has 13 tables and one figure. 
\begin{table}[!ht]
\caption{\label{tab:Numb-MotorRatio}
The number of trajectories used for data sets shown in figures \ref{ContourLength-Displacement} - \ref{Crossing-Turning} where the number of motors on the cargo was changed, and the data was for all mesh sizes. }
\begin{ruledtabular}
\begin{tabular}{cccc}
 Data & 1:1 & 1:5 & 1:10 \\
\hline
chambers & 9  & 7 & 8  \\
movies & 36  & 11 & 15  \\
trajectories & 881  & 270 & 326  \\
\end{tabular}
\end{ruledtabular}
\end{table}

\begin{table}[!ht]
\caption{\label{tab:Fig2Ci}
Best fits for data in figure \ref{ContourLength-Displacement}Ai using equation \ref{eqn:singleExp} or \ref{eqn:doubleExp}, as specified.}
\begin{ruledtabular}
\begin{tabular}{ccccccc}
 data & fit  & $\lambda_1$ ($\mu$m) & $\lambda_2 $  ($\mu$m) & A &$\chi^2$& $R^2$\\
\hline
1:1 & Eq. \ref{eqn:singleExp} & $1.97 \pm 0.02$  & NA & NA & 2.52 & 0.96 \\
1:5 & Eq. \ref{eqn:doubleExp} & $2.5 \pm 0.2$ & $10.0 \pm 0.1$ & $-0.22 \pm 0.02$
& 0.035 & 0.998  \\
1:10& Eq. \ref{eqn:doubleExp} & $1.63 \pm 0.08$ & $10.6 \pm 0.1$ & $-0.23 \pm 0.01$
& 0.039 & 0.999  \\
\end{tabular}
\end{ruledtabular}
\end{table}

\begin{table}[!ht]
\caption{\label{tab:Fig2Cii}
Best fits for data in figure \ref{ContourLength-Displacement}Cii using equation \ref{eqn:singleExp} or \ref{eqn:doubleExp}, as specified.}
\begin{ruledtabular}
\begin{tabular}{ccccccc}
 data & fit  & $\lambda_1$ ($\mu$m) & $\lambda_2 $  ($\mu$m) & A &$\chi^2$& $R^2$\\
\hline
1:1 & Eq. \ref{eqn:singleExp} & $1.73 \pm 0.01$  & NA & NA & 2.83 & 0.954 \\
1:5 & Eq. \ref{eqn:doubleExp} & $1.09 \pm 0.06$ & $8.06 \pm 0.04$ & $-0.177 \pm 0.005$
& 0.0205 & 0.999  \\
1:10& Eq. \ref{eqn:doubleExp} & $1.21 \pm 0.05$ & $8.59 \pm 0.04$ & $-0.260 \pm 0.006$
& 0.035 & 0.999  \\
\end{tabular}
\end{ruledtabular}
\end{table}

\begin{table}[!ht]
\caption{\label{tab:KSFig2A}
Statistical significance comparing the data displayed in figure \ref{ContourLength-Displacement}Ci using the Kolmogorov-Smirnov statistical test.}
\begin{ruledtabular}
\begin{tabular}{cccc}
 Data & 1:1 & 1:5 & 1:10 \\
\hline
1:1 & --  & 0.000 & 0.000  \\
1:5 &   & -- & 0.239  \\
1:10 &   &  & --  \\
\end{tabular}
\end{ruledtabular}
\end{table}

\begin{table}[!ht]
\caption{\label{tab:KSFig2B}
Statistical significance comparing the data displayed in figure \ref{ContourLength-Displacement}Cii using the Kolmogorov-Smirnov statistical test.}
\begin{ruledtabular}
\begin{tabular}{cccc}
 Data & 1:1 & 1:5 & 1:10 \\
\hline
1:1 & --  & 0.000 & 0.000  \\
1:5 &   & -- & 0.096  \\
1:10 &   &  & --  \\
\end{tabular}
\end{ruledtabular}
\end{table}

\begin{table}[!ht]
\caption{\label{tab:Fig3A}
Best fits for data in figure \ref{Duration-Speed-Tortuosity}Ai using equation \ref{eqn:singleExp}.}
\begin{ruledtabular}
\begin{tabular}{ccccc}
 data & fit  & $\lambda$ ($\mu$m) &$\chi^2$& $R^2$\\
\hline
1:1 & Eq. \ref{eqn:singleExp} & $13.4 \pm 0.1$  & 4.42 & 0.929 \\
1:5 & Eq. \ref{eqn:singleExp} & $49.2 \pm 0.3$  & 0.23 & 0.990 \\
1:10& Eq. \ref{eqn:singleExp} & $43.3 \pm 0.4$  & 0.72 & 0.973 \\
\end{tabular}
\end{ruledtabular}
\end{table}

\begin{table}[!ht]
\caption{\label{tab:KSFig3A}
Statistical significance comparing the data displayed in figure \ref{Duration-Speed-Tortuosity}Ai using the Kolmogorov-Smirnov statistical test.}
\begin{ruledtabular}
\begin{tabular}{cccc}
 Data & 1:1 & 1:5 & 1:10 \\
\hline
1:1 & --  & 0.000 & 0.000  \\
1:5 &   & -- & 0.217  \\
1:10 &   &  & --  \\
\end{tabular}
\end{ruledtabular}
\end{table}

\begin{table}[!ht]
\caption{\label{tab:Fig3B}
Best fits for data in figure \ref{Duration-Speed-Tortuosity}Bi using equation \ref{eqn:Gaussian}.}
\begin{ruledtabular}
\begin{tabular}{ccccccc}
 data & fit  & $\mu$ ($\mu$m/s) & $\sigma$ ($\mu$m/s) &$\chi^2$& $R^2$\\
\hline
1:1 & Eq. \ref{eqn:Gaussian} & $0.1468 \pm 0.0003$  & $0.0611 \pm 0.0006$ & 1.26 & 0.980 \\
1:5 & Eq. \ref{eqn:Gaussian} & $0.2788 \pm 0.0004$ & $0.0966 \pm 0.0006$ & 0.087 & 0.996 \\
1:10& Eq. \ref{eqn:Gaussian} & $0.292 \pm 0.0002$ & $0.0983 \pm 0.0003$ & 0.036 & 0.997 \\
\end{tabular}
\end{ruledtabular}
\end{table}

\begin{table}[!ht]
\caption{\label{tab:KSFig3B}
Statistical significance comparing the data displayed in figure \ref{Duration-Speed-Tortuosity}Bi using the Kolmogorov-Smirnov statistical test.}
\begin{ruledtabular}
\begin{tabular}{cccc}
 Data & 1:1 & 1:5 & 1:10 \\
\hline
1:1 & --  & 0.000 & 0.000  \\
1:5 &   & -- & 0.149  \\
1:10 &   &  & --  \\
\end{tabular}
\end{ruledtabular}
\end{table}

\begin{table}[!ht]
\caption{\label{tab:KSFig3C}
Statistical significance comparing the data displayed in figure \ref{Duration-Speed-Tortuosity}Ci using the Kolmogorov-Smirnov statistical test.}
\begin{ruledtabular}
\begin{tabular}{cccc}
 Data & 1:1 & 1:5 & 1:10 \\
\hline
1:1 & --  & 0.080 & 0.386  \\
1:5 &   & -- & 0.748  \\
1:10 &   &  & --  \\
\end{tabular}
\end{ruledtabular}
\end{table}

\begin{table}[!ht]
\caption{\label{tab:Numb-Mesh}
The number of trajectories used for data sets shown in figures \ref{VMeshLengths} and \ref{VMeshTimeSpeedTort} where one quarter of the movie region was used to deduce the average mesh size and the number of trajectories from inside that quadrant are reported. }
\begin{ruledtabular}
\begin{tabular}{ccc}
1:5 & Mesh size ($\mu$m) & Number of Trajectories \\
\hline
 & 0.5 &	22 \\
& 1.4 &	23 \\
& 1.4 &	25 \\
& 1.6 &	22 \\
& 1.7 &	26 \\
& 1.8 &	22 \\
& 1.9 & 24 \\
& 2.2 &	25 \\
& 2.2 &	26 \\
& 2.6 &	22 \\
& 2.8 & 26 \\
\hline
\hline
1:10 & Mesh size ($\mu$m) & Number of Trajectories \\
\hline
& 1.9 &	23 \\
& 2.2 &	15 \\
& 2.2 &	22 \\
& 2.3 &	23 \\
& 2.4 &	19 \\
& 2.5 &	23 \\
& 2.6 &	21 \\
& 2.6 &	23 \\
& 2.6 &	23 \\
& 2.6 &	23 \\
& 2.7 &	19 \\
& 2.8 &	22 \\
& 2.9 &	22 \\
& 3.0 &	23 \\
& 3.1 &	22 \\
\end{tabular}
\end{ruledtabular}
\end{table}

\begin{table}[!ht]
\caption{\label{tab:Fig5B}
Best fits for data in figure \ref{VMeshLengths}B using the equation for a line: $y=mx+b$.}
\begin{ruledtabular}
\begin{tabular}{cccccccc}
 slope (m) & intercept & $\chi^2$& $R^2$\\
\hline
$0.9 \pm 0.7$  & $8 \pm 2$ & 118 & 0.059 \\
\end{tabular}
\end{ruledtabular}
\end{table}

\begin{table}[!ht]
\caption{\label{tab:Fig6}
Best fits for data in figure \ref{VMeshTimeSpeedTort} using the equation for a line: $y=mx+b$.}
\begin{ruledtabular}
\begin{tabular}{cccccccc}
 Data &slope (m) & intercept & $\chi^2$& $R^2$\\
\hline
Fig. \ref{VMeshTimeSpeedTort}A & $-6 \pm 3$  & $54 \pm 7$ & 1710 & 0.142 \\
Fig. \ref{VMeshTimeSpeedTort}B & $0.03 \pm 0.02$  & $0.25 \pm 0.04$ & 0.0516 & 0.116 \\
Fig. \ref{VMeshTimeSpeedTort}C & $-0.2 \pm 0.1$  & $1.9 \pm 0.3$ & 4.46 & 0.0856 \\
masked \\ Fig. \ref{VMeshTimeSpeedTort}C & $-0.16 \pm 0.04$  & $1.6 \pm 0.1$ & 0.224 & 0.390 \\
\end{tabular}
\end{ruledtabular}
\end{table}

\begin{figure}[h!]
\centering
\includegraphics[scale=0.9]{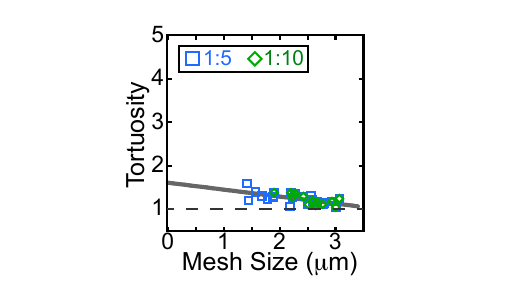}
\caption{{\bf Network mesh size affects the tortuosity.}
Scatter plot of the average tortuosity for quantum dots with 5 motors (blue squares) and 10 motors (green diamonds) as a function of the local network mesh size. Same data presented in figure \ref{VMeshTimeSpeedTort}C with masking and refit with a line. Error bars represent standard error of the mean for the measurements made in the same region with the average network mesh size. Fits given in Appendix Table \ref{tab:Fig6}. The dashed line is the minimum value of  tortuosity, which is one. }
\label{masked}
\end{figure}


\clearpage
\bibliography{biblio}

\end{document}